\documentclass[12pt,preprint]{aastex} 
\usepackage{epstopdf} 

\def\gtorder{\mathrel{\raise.3ex\hbox{$>$}\mkern-14mu
             \lower0.6ex\hbox{$\sim$}}} 
\def\ltsima{$\; \buildrel < \over \sim \;$}
\def\simlt{\lower.5ex\hbox{\ltsima}}
\def\gtsima{$\; \buildrel > \over \sim \;$}
\def\simgt{\lower.5ex\hbox{\gtsima}} 

\begin{document} 


\title{Luminous Supernovae}


\author{Avishay Gal-Yam}
\affil{Department of Particle Physics and Astrophysics, Faculty of Physics, The Weizmann
Institute of Science, Rehovot 76100, Israel}
\email{avishay.gal-yam@weizmann.ac.il}



\begin{abstract} 


Supernovae (SNe), the luminous explosions of stars, were observed since 
antiquity, with typical peak luminosity not exceeding 
$1.2 \times 10^{43}$\,erg\,s$^{-1}$ (absolute magnitude $>-19.5$ mag).  
It is only in the last dozen years that numerous examples
of SNe that are substantially super-luminous ($>7\times10^{43}$\,erg\,s$^{-1}$; 
$<-21$ mag absolute) were well-documented. Reviewing
the accumulated evidence, we define three broad classes of
super-luminous SN events (SLSNe). Hydrogen-rich events (SLSN-II) radiate photons 
diffusing out from thick hydrogen layers where they have been deposited 
by strong shocks, and often show signs of interaction
with circumstellar material. SLSN-R, a rare class of hydrogen-poor events, 
are powered by very large amounts of radioactive $^{56}$Ni and
arguably result from explosions of very massive stars due to the pair instability. 
A third, distinct group of hydrogen-poor events emits photons from 
rapidly-expanding hydrogen-poor material distributed over large radii, and are not 
powered by radioactivity (SLSN-I). These may be the hydrogen-poor analogs of 
SLSN-II. 

\end{abstract} 




\section{Introduction} 

Supernova explosions play important roles in many aspects of astrophysics,
being sources of heavy elements, ionizing radiation and energetic particles;
driving gas outflows and shock waves that shape star and galaxy formation;
and leaving behind compact neutron star and black hole remnants. The study
of supernovae in general has thus been actively pursued for many decades.

\begin{figure}[h]
\centering
\includegraphics[width=1\textwidth]{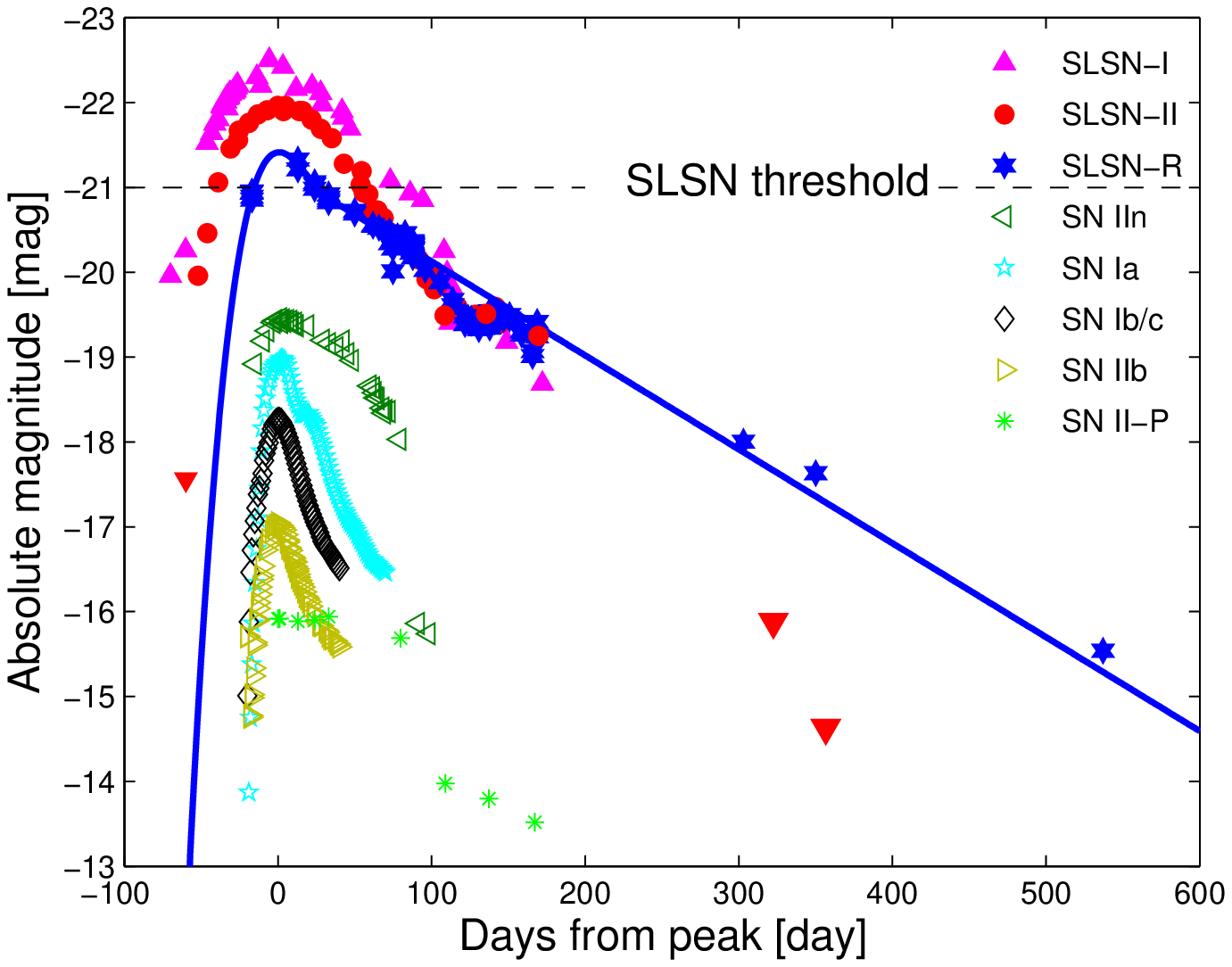}
\caption{The luminosity evolution (light curve) of supernovae. Common SN explosions reach peak luminosities of $\sim10^{43}$\,erg\,s$^{-1}$ 
(absolute magnitude $>-19.5$). The new class of super-luminous SN (SLSN) reach luminosities $\sim10$ times higher. The prototypical 
events of the three SLSN classes (SLSN-I PTF09cnd, Quimby et al. 2011; SLSN-II SN 2006gy, Smith et al. 2007, Ofek et al. 2007, Agnoletto et al. 2009; and
SN 2007bi, Gal-Yam et al. 2009) are compared with a normal Type Ia SN (Nugent template), Type IIn SN 2005cl (Kiewe et al. 2011), the average Type Ib/c 
light curve from Drout et al. (2012), the Type IIb SN 2011dh (Arcavi et al. 2011) and the prototypical Type II-P SN 1999em (Leonard et al. 2002).  
All data are in the observed $R$ band. See SOM for additional details.}
\label{LCfig}
\end{figure}

The discovery of super-luminous supernova (SLSN; Figure~\ref{LCfig}) events in the past decade
is now focusing attention on these extreme explosions. The study of SLSNe
is motivated, among other things, by their likely association with the deaths of
the most massive stars; their potential contribution to the chemical evolution
of the Universe and, at early times, to its reionization; and since they may
be manifestations of physical explosion mechanisms that differ from those of
their more common and less luminous cousins.

With extreme luminosities extending over tens of days (Fig.~\ref{LCfig}) and, in some cases, 
copious ultra-violet (UV) flux, these events may become useful cosmic 
beacons to study distant star-forming galaxies and their gaseous environments;
their long duration, coupled with the lack of long-lasting environmental effects, as well as the fact that they
eventually disappear and allow their hosts to be studied without interference,
offer some advantages over other probes of the distant universe such as 
short-lived gamma-ray burst (GRB) afterglows, and luminous high-redshift Quasars.  
  
Accumulated observations suggest that SLSNe can be naturally grouped into three distinct 
subclasses. In analogy to lower-luminosity explosions, we split SLSNe into hydrogen-rich 
events (SLSN-II) and events lacking spectroscopic signatures of hydrogen (SLSN-I). The latter 
group is further divided into a minority of events whose luminosity is dominated by radioactivity
(SLSN-R) and a distinct majority that require some other source of luminosity, which we simply
refer to as SLSN-I. A detailed discussion of these three classes follows in section $\S~3$ below.  

Supernovae have been traditionally classified mainly according to their
spectroscopic properties (see Filippenko 1997 for a review), 
while the luminosity of SNe does not play a role in the currently used scheme. In principle, 
almost all luminous SNe belong to one of two spectroscopic classes: 
Type IIn, hydrogen-rich events with narrow emission
lines which are usually interpreted as signs of interaction with material lost by the star prior
to the explosion; or Type Ic, events lacking hydrogen, helium, and strong silicon and sulphur lines around
maximum, presumably associated with massive star explosions. However,
as will be discussed below, the physical properties implied by the huge 
luminosities of SLSNe suggest they arise, in many cases, from
progenitor stars that are very different from those of their much more common
and less luminous analogs; we will propose a provisional extension of the
classification scheme that can be applied to super-luminous events. 
 
We consider below SNe with reported peak magnitudes
M$<-21$\,mag {\it in any band} as being superluminous (Fig.~\ref{LCfig}). See Supplementary Online Material (SOM)
section $\S~7$ for considerations related to determining this threshold. 

As will be shown below, modern studies based on large SN samples and homogeneous, CCD-based
luminosity measurements, show that SLSNe are very rare in nearby, luminous and
metal-rich host galaxies (with the possible exception of luminous SNe IIn). One would therefore
expect that older SN surveys, mainly conducted by monitoring nearby galaxies drawn from galaxy catalogs
(dominated by luminous, metal-rich objects, we will refer to these as ``targeted surveys'' for
short) would find very few luminous events. Indeed, 
Richardson et al. (2002) find no solid detections of super-luminous events resulting from the
older targeted surveys (SOM $\S~8$). A handful of events are determined to be 
``genuinely overluminous'' (based on full, modern studies); Richardson
et al. list SN 1997cy and SN 1999as; we will add SN 1999bd for which we present here new
data. These events, found by the first generation of wide-field non-targeted surveys, are indeed
the harbingers of the classes of SLSNe we can finally define today. We note that even the 
brightest SNe associated with cosmological Gamma-Ray Bursts (GRBs, e.g., the nickel-rich SN 1998bw) fall
well below our SLSN threshold.    

The structure of this review is as follows: we begin by a brief summary of the observational work leading to the
discovery of superluminous SNe in section 2, and we synthesize the accumulated information and define 
classes of superluminous SNe in 3. We review open questions and controversies in 4, and conclude with a
summary in 5.

\section{New surveys and the discovery of the first SLSNe}

We now know that superluminous SNe are intrinsically rare. Their discovery therefore
requires surveys that cover a large volume. The first generation of surveys covering
 large volumes were designed
to find numerous distant Type Ia SNe for cosmological use. These observed 
relatively small fields of view to a great depth, placing most of the effective survey volume 
at high redshift. Possible early discoveries of SLSNe by such surveys and their modern
counterparts are discussed in the SOM (section $\S~8$).

\begin{figure}[h]
\centering
\includegraphics[width=1\textwidth]{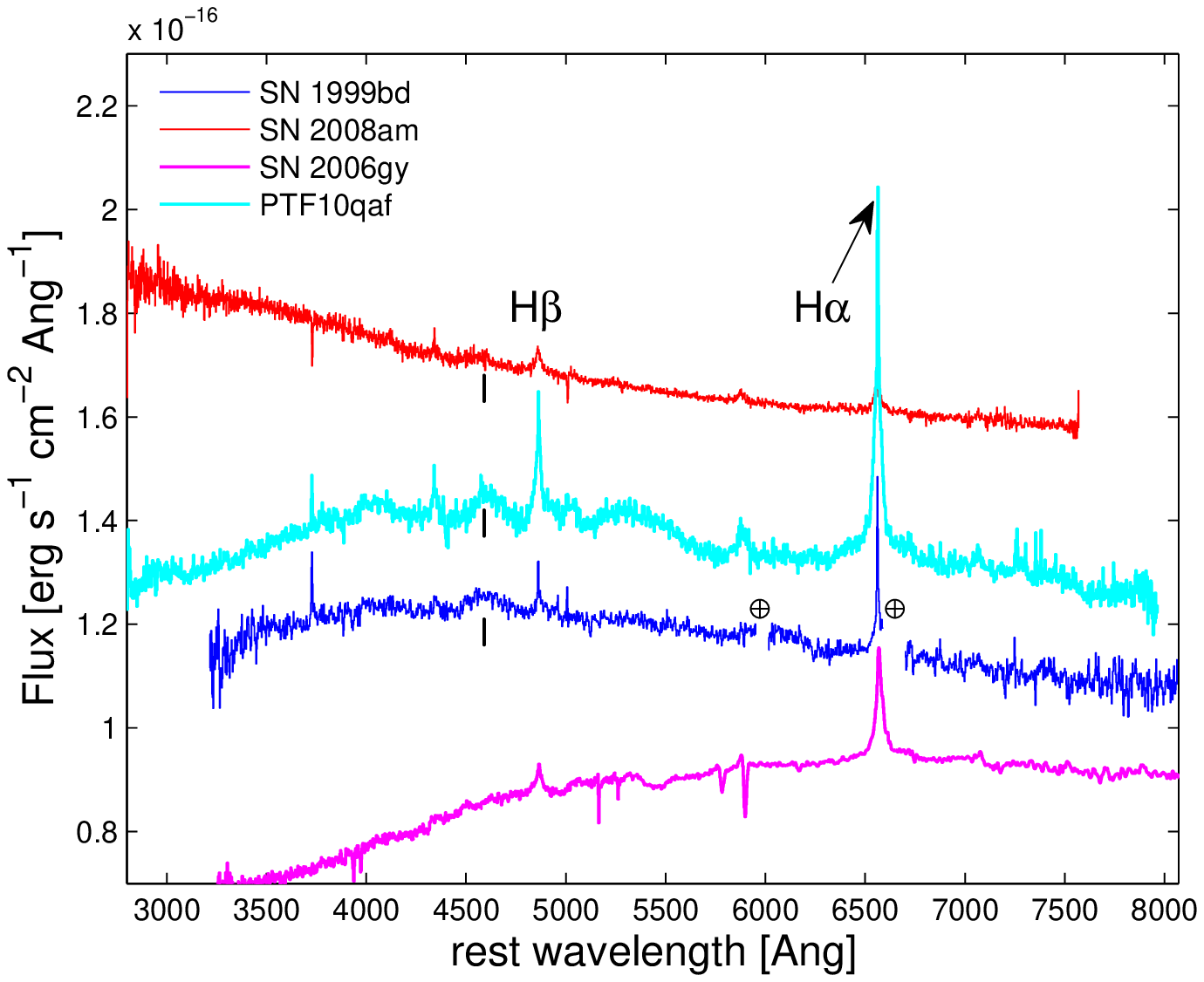}
\caption{Spectra of SLSN-II. A spectrum of SN 1999bd obtained on March 22, 1999 with the 2.5m Dupont telescope at Las Campanas (blue)
is compared with spectra of SN 2006gy (from Smith et al. 2007; magenta), SN 2008am (from Chatzopoulos et al. 2011, red) and a luminous
SLSN-II from PTF (PTF10qaf, cyan). The Balmer lines show narrow and intermediate-width
components (compare with the narrow host oxygen [OII] emission lines). A prominent emission bump around $4600$\,\AA (black tick marks) 
is also a common feature. At a redshift of z=0.1512, the absolute magnitude
at discovery of SN 1999bd was -21.6. Telluric bands are marked and sections of the spectrum of SN 1999bd affected have been excised; 
the telluric A band strongly absorbs the red wing of the H$\alpha$ line in this spectrum.  
}
\label{sn1999bd-spec}
\end{figure}

An alternative method to survey a large volume is to use wide-field facilities
to cover a large sky area with relatively shallow imaging. With most of the survey 
volume at low redshift one can conduct an efficient untargeted survey for 
nearby SNe. Such surveys provided the first well-observed examples of SLSNe. Prominent initial
discoveries were made by the cosmology group at Lawrence Berkeley Labs (Regnault et al. 2001), 
e.g., SN 1999as (Knop et al. 1999), which turned out to be
the first example of the class of extremely $^{56}$Ni-rich Type I SNe (Gal-Yam et al. 2009) 
and SN 1999bd (Nugent et al. 1999; Fig.~\ref{sn1999bd-spec}), which is probably
the first well-documented case of a superluminous SN IIn. The study of the peculiar SN 1997cy by   
Germany et al. (2000) is perhaps the first modern study of a supernova which
is substantially more luminous than the norm, though the absolute peak magnitude of this event (-20.1 mag) as well as those of similar objects discovered since (e.g., SN 2002ic, Hamuy et al. 2003) fall below our fiducial limit defined above, and we do not discuss them any further.

Perhaps the most significant breakthrough in our observational knowledge about the most luminous SN explosions resulted from the
operation of the Texas Supernova Survey (TSS) by Quimby and collaborators (Quimby 2006). 
On 2005 March 3 this survey discovered a hostless transient at mag 18.13, its redshift was $z=0.2832$, indicating an absolute magnitude at peak around $-22.7$\,mag, marking 
it as the most luminous SN discovered until that time (Quimby et al. 2007). SN 2005ap is the first example of the class we will define below as SLSN-I.   
On 2006 September 18 TSS discovered a bright transient located at the nuclear region of the nearby galaxy NGC 1260 (SN 2006gy; Smith et al. 2007). 
SN 2006gy suffered substantial host extinction; correcting for this effect, the measured peak magnitude was extreme ($\sim-22$\,mag; Ofek et al. 2007; Smith et al. 2007). Spectroscopy of SN 2006gy 
clearly showed hydrogen emission lines with both narrow and intermediate-width components, leading
to a spectroscopic classification of SN IIn; this is the prototype and best-studied example of the SLSN-II class.   
TSS went on to discover additional luminous SNe, as described in the SOM.

During the last few years, several untargeted surveys have been operating in parallel, and now discover and report most SN events  (Gal-Yam \& Mazzali 2010).
The large volume probed by these surveys and their coverage of a multitude of low-luminosity dwarf galaxies have led, as expected (Young et al.
2008) to the discovery of numerous unusual SNe not seen before in targeted surveys of luminous hosts; indeed, it was shown that the SN population
in dwarf galaxies is different than that observed in giant hosts (Arcavi et al. 2010) . As will be discussed below, luminous SNe generally tend
to prefer dwarf hosts (e.g., Neill et al. 2011) and indeed numerous such events are being discovered by this new generation of untargeted 
surveys. The Catalina Real-Time Transient Survey (CRTS) reported numerous SLSN events including SN 2008fz,
an extremely luminous member of the SNLS-II family (Drake et al. 2010). Results from the Palomar Transient Factory (PTF; Law et al. 2009, Rau et al. 2009)
include the discovery of SN 2007bi (Gal-Yam et al. 2009), 
the first extensively-observed example of  the radioactively-powered superluminous SN variety (SLSN-R) during a ``dry-run'' initial phase of the
project in 2007, while the more recent PTF discovery of several members of the class of SLSN-I allowed Quimby et al. (2011) to define this
class and derive its main properties. The Panoramic Survey Telescope and Rapid Response System 1 (Pan STARRS 1; hereafter PS1, Kaiser et al. 2010) appears to be an efficient program to detect numerous SLSNe at high redshifts (out to
$z\sim1$ and beyond, e.g., Chomiuk et al. 2011; Pastorello et al. 2010). More details about these surveys and their results are provided in the SOM.    

\section{Emerging classes of superluminous SNe}

A total of 18 SLSNe have been discussed in the literature (Table 1). These objects can be grouped into
three classes that share observational and physical attributes. We now describe each of these classes in detail. 

\subsection{SLSN-R}

Of all classes of super-luminous SNe, this seems to be the best understood. SLSN-R events are powered by large amounts (several M$_{\odot}$) 
of radioactive $^{56}$Ni (hence the suffix ``R''), produced during the explosion of a very massive star. The radioactive decay chain  $^{56}$Ni$\rightarrow^{56}$Co$\rightarrow ^{56}$Fe deposits energy via $\gamma$-ray and positron emission, that is thermalized and converted to optical radiation by the expanding massive ejecta. The luminosity of the peak is broadly proportional to the amount of radioactive $^{56}$Ni, while the late-time decay (which in the most luminous
cases begins immediately after the optical peak) follows the theoretical $^{56}$Co decay rate ($0.0098$\,mag\,day$^{-1}$). The luminosity of this
``cobalt radioactive tail'' can be used to infer an independent estimate of the initial $^{56}$Ni mass.

The first well-observed example of this group was SN 2007bi, discovered by the PTF ``dry run'' experiment. An extensive investigation of this object and its physical nature is presented in Gal-Yam et al. (2009). The most prominent physical characteristic of this group, the large $^{56}$Ni mass, is well-measured in this case using both the peak luminosity ($R=-21.3$\,mag) and the cobalt decay tail, followed for $>500$ days. Estimates derived from the observations, as
well as via comparison to other well-studied events (SN 1987A and SN 1998bw) converge on a value of  M$_{^{56}{\rm Ni}}\approx5$\,M$_{\odot}$.
The large amount of radioactive material powers a long-lasting phase of nebular emission, during which the optically thin ejecta are energized by the 
decaying radio nucleides. Analysis of late-time spectra obtained during this phase (Gal-Yam et al. 2009) provides independent confirmation of the large 
initial $^{56}$Ni mass via detection of strong nebular emission from the large mass of resulting $^{56}$Fe, as well as the integrated emission from
all elements, powered by the remaining $^{56}$Co. 

Estimation of other physical parameters of the event, in particular the total ejected mass (which provides a lower limit on the progenitor star mass), its 
composition, and the kinetic energy it carries, is more complicated. There are no observed signatures of hydrogen in this event (either in the ejecta or
traces of CSM interaction) so the ejecta mass directly constrains the mass of the exploding helium core, which is likely dominated by oxygen and heavier elements. Gal-Yam et al. (2009) use scaling relations based on the work of Arnett (1982), as well as comparison of the data to custom light-curve models, and derive an ejecta mass of M\,$\approx100$\,M$_{\odot}$. Analysis of the nebular spectra provides an independent lower limit on the mass of  M\,$>50$\,M$_{\odot}$,
with a composition similar to that expected from theoretical models of massive cores exploding via the pair-instability process.
Moriya et al. (2010) postulate a lower ejecta mass (M\,$=43$\,M$_{\odot}$); this difference becomes crucial to the controversy about the explosion mechanism of these giant cores (see below). In any case there is no doubt these explosions are produced by extremely massive stars, with the most 
massive exploding heavy-element cores we know. The same scaling relations used by Gal-Yam et al. (2009) also indicate extreme values of ejecta kinetic
energy (approaching E$_k=10^{53}$\,erg). Finally, the integrated radiated energy of this event over its very long lifetime is high ($>10^{51}$\,erg).       
   
\begin{figure}[h]
\centering
\includegraphics[width=1\textwidth]{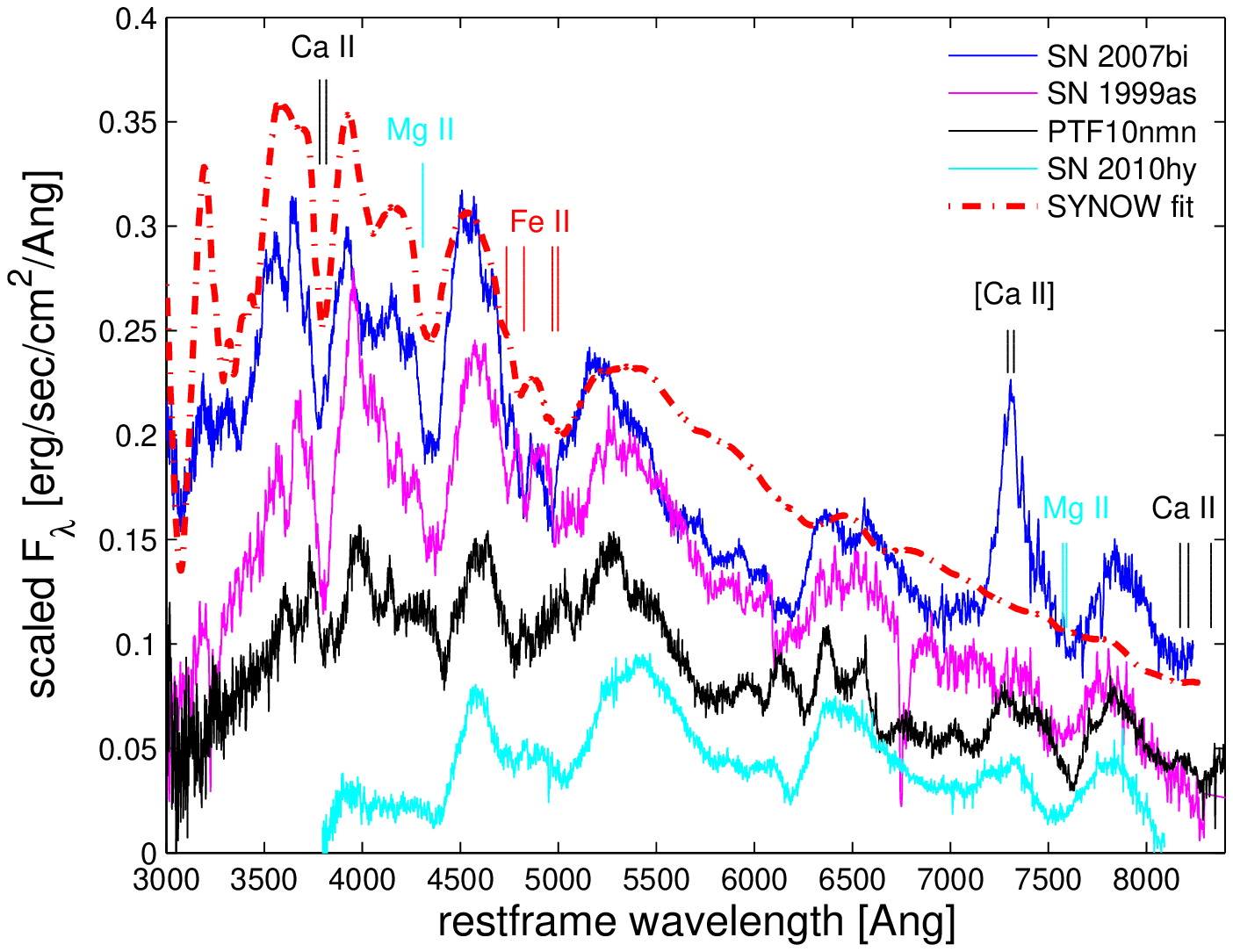}
\caption{Photospheric spectra of SLSN-R events SN 2007bi (blue, from Gal-Yam et al. 2009), SN 1999as (magenta, from Gal-Yam et al. 2009),  PTF10nmn (black, from Gal-Yam 2011) and SN 2010hy (cyan; S. B. Cenko, private communication); all spectra were obtained close to peak. 
Identification of prominent spectral features as well as a synthetic SYNOW fit (red, 
from Gal-Yam et al. 2009) are also shown.
}
\label{PISNfig}
\end{figure}

SN 1999as was one of the first genuine SLSNe discovered (Knop et al. 1999, see above) and was initially analyzed by Hatano et al. (2001). 
As shown in Gal-Yam et al. (2009; see also Fig.~\ref{PISNfig}) 
this object was similar to SN 2007bi during its photospheric phase
(reaching $-21.4$\,mag absolute at peak),
while the analysis of Hatano et al. (2001) suggests physical attributes ($^{56}$Ni mass, kinetic energy, and ejected mass) that are close to, but somewhat
lower than, those of SN 2007bi. Unfortunately, no late-time data have been collected for this object, so it is impossible to conduct the same analysis 
carried for SN 2007bi, but the similarities suggest this was another member of the SLSN-R class. 

Recently, the Lick Observatory Supernova Survey (LOSS; Filippenko et al. 2001) using the 0.75m Katzman Automatic Imaging Telescope (KAIT) discovered the luminous Type Ic SN 2010hy (Kodros et al. 2010; Vinko et al. 2010); Following the discovery by KAIT this event was also recovered in PTF data 
(and designated PTF10vwg). It is interesting to note that while the LOSS survey is operating in a targeted mode looking at a list of known galaxies,
by performing image subtraction on the entire KAIT field of view it is effectively running in parallel also an untargeted survey of the background
galaxy population (as noted by Gal-Yam et al. 2008 and Li et al. 2011). It is during this parallel survey that KAIT detected this interesting rare SN,
residing in an anonymous dwarf host. While
final photometry is not yet available for this event, preliminary KAIT and PTF data indicate a peak magnitude of $-21$\,mag or brighter, and it is 
spectroscopically similar to other SLSN-R (Fig.~\ref{PISNfig}) suggesting it is also likely a member of this class.
      
Objects of this sub-class are exceedingly rare (this is observationally the rarest class among the SLSN classes reviewed here; I discuss relative
volumetric rates below), and thus additional examples are scarce. During the last two years, the PTF survey has discovered another likely
member (PTF10nmn; Gal-Yam 2011; Fig.~\ref{PISNfig}) with similar properties to SN 2007bi, while PS1 may have discovered another similar object at a higher redshift (Smartt 2011). Assembling a reasonable sample of such events may thus be a time-consuming process. 

Young et al. (2010) present additional observations of SN 2007bi during the decline phase, as well as a detailed study
of its host galaxy. They find the host is a dwarf galaxy (with luminosity similar to that of the SMC), with relatively low 
metallicity ($Z\approx Z_{\odot}/3$) - somewhere between those of the LMC and SMC. So, while the progenitor star of this explosion probably had sub-solar metal content, there
is no evidence that it had very low metallicity. The host galaxy
of SN 1999as is more luminous (and thus likely more metal-rich) than that of SN 2007bi, but still fainter than typical giant galaxies like the Milky Way (Neill et al. 2011), while the host of PTF10nmn seems to be
as faint or fainter than that of SN 2007bi. It thus seems this class of objects typically explode in dwarf galaxies.

Considering all available data, it seems there is agreement about the observational properties of this class and their basic interpretation: very massive star explosions that produce large quantities of radioactive $^{56}$Ni. A controversy still exists about the underlying explosion mechanism that leads to
this result, either very massive oxygen cores (M$>50$\,M$_{\odot}$) become unstable to electron-positron pair production and collapse (Gal-Yam et al. 
2009), or else slightly less massive cores (M$<45$\,M$_{\odot}$) evolve all the way till the common iron-core-collapse process occurs (Moriya et al. 2010). 
We will return to this question below.
Assuming, for the sake of the current discussion, that these explosions do arise from the pair instability, a clear prediction of the relevant theoretical
models (e.g., Heger \& Woosley 2002, Waldman 2008) is that for each luminous, $^{56}$Ni-rich explosion (from a core around $100$\,M$_{\odot}$) there
would be numerous less luminous events with smaller $^{56}$Ni masses but large ejecta masses (M$>50$\,M$_{\odot}$). These should manifest
as events with very slow light curves (long rise and decay times) and yet moderate or even low peak luminosities (SOM; Fig.~\ref{PISN-stats}). 

\subsection{SLSN-II}

This is probably the most commonly observed class of SLSN. While some examples have been identified early on
(e.g., SN 1999bd discussed above; Fig.~\ref{sn1999bd-spec}), these objects only became a focus of attention with the discovery of 
SN 2006gy (Ofek et al. 2007; Smith et al. 2007). Since then, several additional examples have been studied in some detail. 
SLSN-II show strong hydrogen features in their spectra; these explosions therefore typically
occur within thick hydrogen envelopes, making investigations of their nature more complicated, as all information carried by 
electromagnetic radiation from the exploding core is reprocessed by the outer envelope. For this reason, our knowledge about 
the energy source (or sources) of these explosions is still mostly speculative. On the other hand, the physics responsible for 
converting the explosion energy into the observed radiation is better understood.

Two main physical processes have been invoked to explain the conversion of explosion energy to emitted radiation in SLSN-II. The
first process assumes the explosion launches a powerful shock wave expanding outwards from the center of the star; this shock heats
the material it traverses until it eventually escapes from the {\it effective} outer edge of the star, where this effective edge is the
radius around which the material in no longer optically thick to radiation. Note that this effective edge does not necessarily coincide
with the physical edge of the star, which we can define here as the radius inside which material is gravitationally bound to the star. 
The energy deposited by the shock is then slowly re-emitted by the hydrogen-rich 
material as photons diffuse out, in analogy to the more common and much less luminous 
Type II-P SNe where this process occurs within the envelope of a red supergiant star. In order to achieve the much higher luminosities 
observed for SLSN-II, the radius of the effective edge of the star must be substantially larger 
compared to radii of even the largest red supergiants; deposition of
shock energy into more compact stars is radiatively inefficient as the deposited energy is quickly drained by adiabatic expansion. The
observed luminosities probably require also an energetic explosion shock. The shape of the light curve is determined mostly
by the density structure and composition of the material into which the energy was deposited. In order to explain the large effective 
radii ($>10^{15}$\,cm) required for this mechanism to work, several options have been suggested. These include very large (bloated) stars 
(e.g., Gezari et al. 2009), energy deposited into massive (unbound) optically-thick shells ejected by previous eruptions of the exploding star
that have expanded to the required radius (e.g., Smith \& McCray 2007; Miller et al. 2009) 
or energy deposited into an optically-thick massive stellar wind
(e.g., Ofek et al. 2010; Chevalier \& Irwin 2011; Moriya \& Tominaga 2011) extending out to the required radius.    

The second mechanism invoked in order to convert large explosion energies into optical emission is strong interaction between the expanding 
ejecta and massive circum-stellar material (CSM), previously lost from the progenitor star. This mechanism converts the kinetic energy carried
by the expanding ejecta into radiation via strong shocks, and is commonly invoked for Type IIn SNe. Note that there is no conflict between this
process (converting kinetic energy to radiation) and the previous one (converting shock energy stored as internal heat energy into radiation)
and both can contribute in any given object; there are some debates in the literature which is dominant for particular cases. Since CSM envelopes 
can be extremely extended, this process can in principle remain active for many years, and is thus more attractive to explain very long-lived
events (e.g., SN 2006tf, Smith et al. 2008; SN 2003ma, Rest et al. 2011). On the other hand, conversion of kinetic energy into radiation
should manifest in an observed decline in expansion velocities; this process is therefore disfavored for events showing high expansion velocities
that do not decrease significantly with time (e.g., SN 2008es, Gezari et al. 2009). Possible mechanisms invoked to eject large quantities of
mass from the star prior to explosions include LBV-like activity (e.g., Smith et al. 2007; 2008) and the pulsational pair instability (e.g. Woosley 
et al. 2007).

Regardless of the conversion mechanism, the total emitted energy in several recently observed objects ($>10^{51}$\,erg) is challenging
to produce in regular iron-core-collapse explosions (where $>99\%$ of the initial explosion energy, $\sim3\times10^{53}$\,erg, 
is carried away by neutrinos). This led several
authors to speculate about additional energy sources contributing to these powerful explosions. Pair instability explosions can provide 
large kinetic energies and synthesize large amounts of radioactive $^{56}$Ni; however, SLSN-II studied at late time did not follow the
expected $^{56}$Co radioactive decay rate, in contrast to SLSN-R; the derived limits on the amount of initial radioactive nickel generally
argue against SLSN-II resulting from energetic pair-instability explosions. Spin-down of nascent magnetars (rapidly spinning neutron stars with
strong magnetic fields) has been proposed as an alternative energy source (Woosley 2010, Kasen \& Bildsten 2010), this process
may be relevant at least for some SLSN-II (e.g., SN 2008es). One can also consider the so-called ``collapsar'' scenario where energy 
is extracted from material rapidly accreting onto a newly-formed black hole; this process may be driving cosmological $\gamma$-ray bursts
(Macfadyen \& Woosley 1999). When occurring within a massive star with a thick hydrogen envelope, this process may deposit the energy
in the expanding envelope, where it may be thermalized and reemitted in the optical (Young et al. 2005; Quimby et al. 2007). Unfortunately,
the fact that any energy injected by such processes at the deep layers of the exploding stars is then reprocessed by the outer, optically
thick hydrogen layers, makes investigations of such exotic processes in SLSN-II difficult, and mostly speculative so far. 

In any case, it is 
clear that SLSN-II are explosions of massive stars that retained their hydrogen envelopes until they exploded. For some objects spectroscopy
indicates these stars have lost substantial amounts of mass prior to explosion (e.g., SN 2006gy, Smith et al. 2007; SN 2006tf, Smith et al. 2008), 
suggesting perhaps the progenitor stars are similar to massive luminous blue variables (LBVs) which are known to undergo
episodic eruptions involving extreme mass loss (e.g., Humphreys \& Davidson 1994); a similar relation has been established in the case
of a lower-luminosity SN IIn (SN 2005gl, Gal-Yam et al. 2007a; Gal-Yam \& Leonard 2009).  

The observational characteristics of SLSN-II are quite diverse. The peak luminosities of the brightest events are impressive, reaching well
above $-22$\,mag absolute (e.g., SN 2008fz, Drake et al. 2010). However, these seem to be the top of a broad distribution, with examples 
of luminous Type IIn events reported with peak magnitudes smoothly extending from these extreme values down to luminosities typical of
the general SN population (e.g., SN 2010jl, Stoll et al. 2011, Smith et al. 2011; see Kiewe et al. 2011 for a review of older events). The light
curve shapes are quite diverse, with some SLSN-II showing a rapid rise and decline (e.g., SN 2008es, Gezari et al. 2009), some
showing light curves with a slow rise ($>50$ days) to a broad peak (e.g., SN 2006gy, Smith et al. 2007, Ofek et al. 2007; SN 2008fz, 
Drake et al. 2010), and some with rapid rise and very slow decline (e.g., SN 2003ma, Rest et al. 2011; SN 2008am, Chatzopoulos et al. 2011; 
and probably also SN 2006tf, Smith et al. 2008). Spectra (Fig.~\ref{sn1999bd-spec})
also show diversity, with most objects showing narrow hydrogen Balmer lines, and SN 2008es uniquely not. Narrow Balmer lines 
arise from a slow wind blown by the progenitor star prior to its explosion, assumed to have been photoionized by the explosion and to then recombine.
The lack of such narrow lines in the spectra of SN 2008es suggests the progenitor star was not blowing massive winds for an extended period prior
to its explosion; any substantial mass loss must have been episodic (e.g., Miller et al. 2009) or otherwise time-variable (Moriya et al. 2011). 

The environments of SLSN-II are also quite diverse. This is the only SLSN subclass that has been detected also in luminous, Milky-Way-like
galaxies (e.g., SN 2006gy, Ofek et al. 2007, Smith et al. 2007; CSS100217:102913+404220, Drake et al. 2011). Still, like other SLSNe,
most SLSN-II events reside in dwarf star-forming hosts (Neill et al. 2011). It is interesting to note that published SLSN-II events in giant
host galaxies seem to have been discovered very close to their host nuclei, suggesting that perhaps specific conditions that are unique to
this environment (e.g., in circumnuclear star-forming rings) somehow mimic the conditions in star-forming dwarf galaxies. A review of 
SLSN-II events from the PTF survey confirms that five additional unpublished events reside either in dwarf hosts or in nuclei of giant
hosts, supporting this interesting trend. Since several SLSN-II have exploded in the high-metallicity cores of luminous galaxies, it seems
that these events do not require low-metallcity progenitors.   

\subsection{SLSN-I}

\begin{figure}[h]
\centering
\includegraphics[width=1\textwidth]{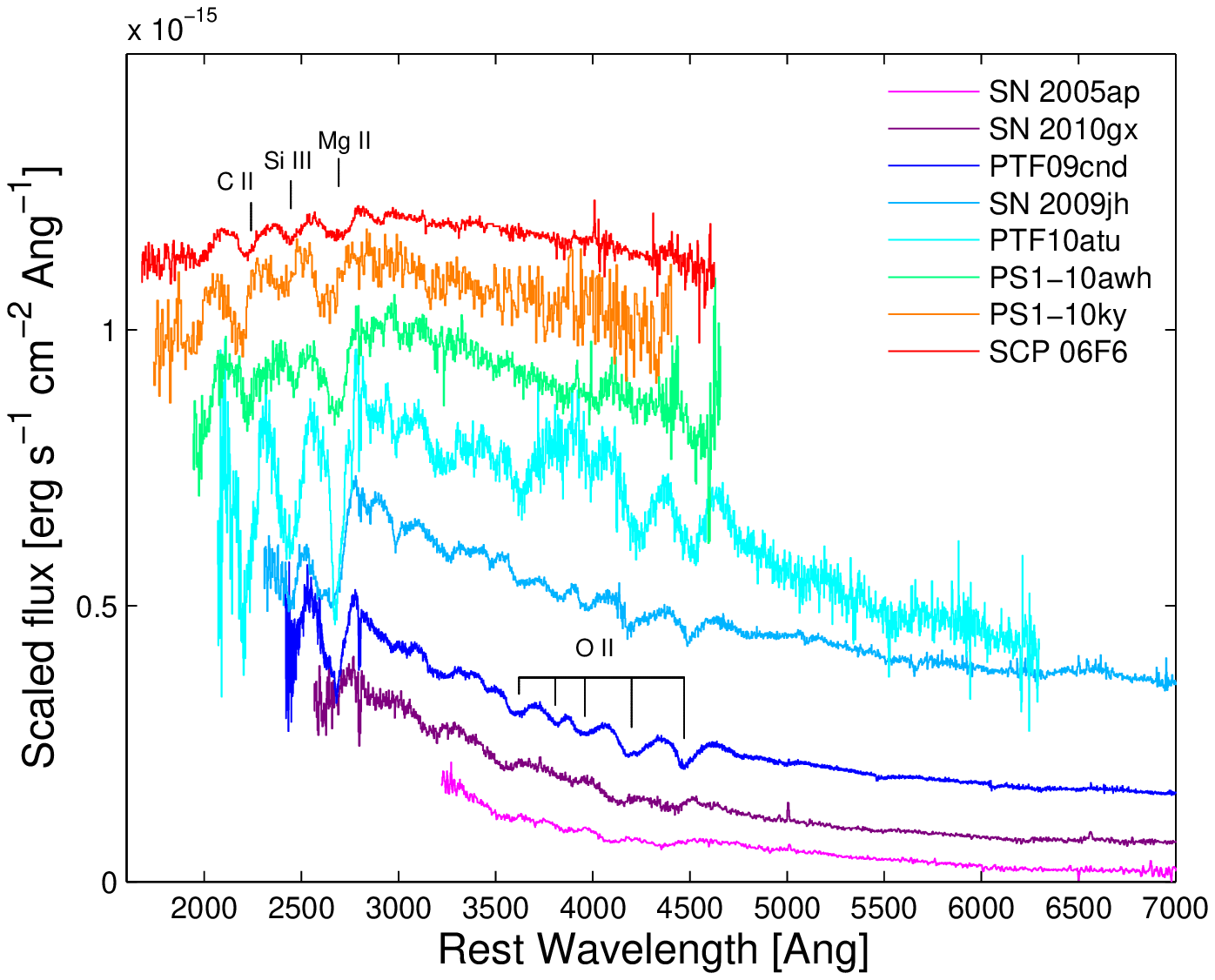}
\caption{Early spectra of all published SLSN-I events: SN2005ap (Quimby et al. 2007); SN 2010gx, PTF09cnd, SN 2009jh, PTF09atu (Quimby et al. 2011); 
PS1-10awh, PS1-10ky (Chomiuk et al. 2011) and SCP 06F6 (Barbary et al. 2009; combined version from Quimby et al. 2011). The optical OII
blends and near-UV CII, Si III  and Mg II lines identified by Quimby et al. (2011) are marked. The spectroscopic similarity among these
objects is quite striking. All spectra shown are available in digital form from the Weizmann Institute of Science Experimental Astrophysics Spectroscopy System (WISEASS) at http://www.weizmann.ac.il/astrophysics/wiseass/).   
}
\label{SLSN-I-spec}
\end{figure}

This final group of SLSNe was initially the most difficult to understand, with the two first reported events, SN 2005ap (Quimby et al. 2007) and
SCP 06F6 (Barbary et al. 2009) rather sparsly observed. It was only following the discovery of additional members of this class by the PTF
survey, bridging the redshift gap between the relatively nearby SN 2005ap ($z=0.2832$) and the high-redshift SCP 06F6 ($z=1.189$) that a comprehensive
view of this class of objects could be formed (Quimby et al. 2011; including spectroscopic redshifts based on MgII absorption lines, and the first correct identification of the redshift of SCP 06F6; Fig.~\ref{SLSN-I-spec}). A major
reason for the initial difficulties was that early spectra of these objects are quite featureless and the absorption lines that do appear are mostly
of high-excitation low-mass elements (Fig.~\ref{SLSN-I-spec}), while the elements commonly observed in most SN classes (neutral oxygen, magnesium, iron and the ubiquitous ionized calcium) appear only much later (Pastorello et al. 2010; SOM Fig.~\ref{sn2005ap-SYNOW}).  

Observationally, these events are characterized by extreme peak luminosities (often brighter than $-22$\,mag absolute), very blue spectra with
significant ultra-violet (UV) flux persisting for many weeks, and (compared to other classes of SLSNe) relatively fast-evolving light curves with
rise times below 50 days and post-peak slopes that decline substantially faster than radioactive cobalt decay rates (Pastorello et al. 2010;
Quimby et al. 2011). Contrary to early
reports (e.g., Quimby et al. 2007, Miller et al. 2009, Gezari et al. 2009), these events do not show hydrogen in their spectra (Quimby et al. 2011;
Pastorello et al. 2010; SOM Fig.~\ref{sn2005ap-SYNOW}), and thus do not belong to the spectroscopic class of Type II SNe. Technically, these events
should be classified as Type Ic SNe (as they also do not show strong He features in spectra taken around peak); since the class of 
Type Ic SNe is not positively defined and a physical connection between the events we consider here and more common SN Ic events has not
been firmly established (see below), we refer to this class as SLSN-I. 
 
Quimby et al. (2011) provide extensive discussion about the physical nature of these objects based on the accumulated observational data. They
show that the observed luminosity requires the deposition of significant amount of internal energy, taking place at large radii 
($10^{15}$\,cm, approximately ten times the size of the largest red supergiants),
 into material expanding at high velocities ($10^4$\,km\,s$^{-1}$). The data can rule out traditional mechanisms discussed above
(radioactivity; photon diffusion, interaction with massive hydrogen-rich CSM). Viable options for the energy conversion mechanism include 
interaction with expanding shells of hydrogen-free material (Chevalier \& Irwin 2011), perhaps ejected by the pulsational pair-instability 
(e.g., Woosley et al. 2007); or reemission of energy injected by an internal engine, such as magnetar spin-down (Woosley 2010; Kasen \& Bildsten
2010) or a ``collapsar''-like accreting black hole (e.g., Quimby et al. 2007); further discussed by Pastorello et al. (2010), Chomiuk et al. (2011)
and Leloudas et al. (2012). Frustratingly, at this time, and even though these events are not
enshrouded by massive, opaque hydrogen shells, the physical nature of the energy source remains speculative, and the energy conversion
mechanism is also not clearly understood.

The host galaxies of these events are again typically dwarf galaxies, though at higher redshifts luminosity upper limits on undetected hosts
are less constraining (Neill et al. 2011, Quimby et al. 2011). The most natural explanation for these objects not occurring in more luminous galaxies 
is that a lower metallicity is required to form the progenitor stars of these events, but other explanations are also possible (e.g., different 
star formation modes or a top-heavy IMF in dwarf galaxies). The extreme intrinsic luminosity and plentiful UV flux of these sources make 
them ideal probes of dwarf galaxies at high redshifts; prospects for detecting numerous such events by deep surveys such as the PS1
medium-deep survey (Chomiuk et al. 2011) seem bright.  

\begin{table}[h] 
\centering 
\begin{tabular}{lllll} 
\hline
\hline 
Supernova & Redshift & Absolute       & Radiated                      & Reference\\
                  &              & peak [mag]     & energy [erg]                &                 \\
\hline 
SLSN-R      &               &                       &                                     &                                \\
SN 2007bi  & 0.1289  & -21.3              & $1-2 \times 10^{51}$ & Gal-Yam et al. 2009\\ 
SN 1999as & 0.12      & -21.4              &                                     & Hatano et al. 2001\\ 
\hline
SLSN-II       &              &                        &                                    &                                \\
CSS100217 & 0.147   & -23.0              & $1.3 \times 10^{52}$  & Drake et al. 2011   \\
SN 2008fz   & 0.133   & -22.3              & $1.4 \times 10^{51}$  & Drake et al. 2010 \\
SN 2008am & 0.2338 & -22.3              & $2 \times 10^{51}$    & Chatzopoulos et al. 2011 \\
SN 2008es  & 0.205   & -22.2              & $1.1 \times 10^{51}$  & Gezari et al. 2009; Miller et al. 2009 \\
SN 2006gy  & 0.019   & -22.0              & $2.3-2.5 \times 10^{51}$ & Ofek et al. 2007; Smith et al. 2010 \\
SN 2003ma & 0.289    & -21.5              &  $4 \times 10^{51}$    & Rest et al. 2011       \\        
SN 2006tf   & 0.074   & -21.0               &$7 \times 10^{50}$      & Smith et al. 2008 \\
\hline
SLSN-I         &              &                       &                                     &                                 \\
SN 2005ap  & 0.2832  & -22.7             & $1.2 \times 10^{51}$   & Quimby et al. 2007; 2011 \\
SCP 06F6    & 1.189    & -22.5             & $1.7 \times 10^{51}$   & Quimby et al. 2011 \\
PS1-10ky     & 0.956   & -22.5              & $0.9-1.4 \times 10^{51}$ & Chomiuk et al. 2011 \\   
PS1-10awh  & 0.908   & -22.5              & $0.9-1.4 \times 10^{51}$ & Chomiuk et al. 2011 \\   
PTF09atu     & 0.501   & -22.0              &                                     & Quimby et al. 2011 \\   
PTF09cnd     & 0.258   & -22.0              & $1.2 \times 10^{51}$ & Quimby et al. 2011 \\   
SN 2009jh    & 0.349   & -22.0              &                                     & Quimby et al. 2011 \\   
SN 2006oz   & 0.376   & -21.5              &                                     & Leloudas et al. 2012 \\
SN 2010gx   & 0.230   & -21.2              & $ 6 \times 10^{50}$   & Quimby et al. 2011; Pastorello et al. 2010 \\   
\hline 
\end{tabular} 
\caption{SLSN properties. We list the reported redshifts, absolute peak magnitudes and total radiated energies
as taken from the literature. The published post-peak magnitude of SN 2006tf (Smith et al. 2008; M$<-20.7$\,mag)
is below our fiducial cutoff. However, Smith et al. argue it was probably brighter at peak; this is confirmed by unpublished TSS 
data that do cover the peak which is around $-21$\,mag absolute (R. Quimby, private communication).} 
\end{table}

\section{Open questions and controversies}

\subsection{The explosion mechanism of SLSN-R: pair instability vs. iron-core-collapse}

While the observations of SLSN-R strongly indicate these events are powered by massive-star explosions that synthesize several
solar masses of radioactive $^{56}$Ni, the physical nature of the explosion is a matter of some controversy. Theoretical works
suggest two options. The first is an extreme version of the iron-core-collapse model that is generally assumed to take place in explosions
of massive stars that manifest as common Type II SNe (Umeda \& Nomoto 2008; Moriya et al. 2010). 
The second is the pair-instability mechanism (e.g., Rakavy \& Shaviv 1967;
Barkat, Rakavy \& Sack 1967; Bond et al. 1984; Heger \& Woosley 2002; Scannapieco et al. 2005; Waldman 2008). The pair 
instability occurs during the evolution of very massive stars that develop  oxygen cores above a critical mass threshold ($\sim50$\,M$_{\odot}$);
these cores achieve high temperatures at relatively low densities; significant amounts of electron-positron pairs are created prior
to oxygen ignition; loss of pressure support, rapid contraction, and explosive oxygen ignition follow, leading to a powerful explosion that
disrupts the star. Extensive theoretical work indicates this result is unavoidable for massive oxygen cores; when the core mass in 
question is large enough ($\sim100$\,M$_{\odot}$, as inferred for SN 2007bi) many solar masses of radioactive nickel are
naturally produced.  

Umeda et al. (2008) and Moriya et al. (2010) show that if one considers a 
carbon-oxygen core with a mass of $\sim43$\,M$_{\odot}$ (just below the pair-instability threshold), 
which explodes with an ad-hoc large explosion energy ($>10^{52}$\,erg),
one can produce the required large amounts of nickel (Umeda \& Nomoto 2008), as well as recover the light curve shape of the
SLSN-R prototype, SN 2007bi (Moriya et al. 2010). Since both the pair-instability model and the massive core-collapse 
model fit the light curve shape of SN 2007bi equally well; and progenitors of pair-instability explosions have larger cores and thus
larger initial stellar masses, which are, assuming a declining initial mass function, intrinsically more rare, Yoshida \& Umeda (2011)
favor the core-collapse model. 

The two models (pair instability vs. core-collapse) agree about the nickel mass, but strongly differ in their predictions about
the {\it total} ejected mass. Total heavy-element masses above the $50$\,M$_{\odot}$ threshold would indicate a core that
is bound to become pair unstable, and will rule out the core-collapse model. Gal-Yam et al. (2009) estimated the total ejected 
heavy-element mass of SN 2007bi in several ways, including modelling of the nebular spectra of this event.  
The core-collapse model of Moriya et al. 2010 does not fit these data (Gal-Yam 2011); this model assumes
a similar amount of radioactive $^{56}$Ni and lower total ejected mass (to avoid the pair-instability) leading to very strong
nebular emission lines that are not consistent with the data. Thus this model is not
viable for this prototypical SLSN-R object, supporting instead a pair instability explosion as originally claimed. 
It remains to be seen whether the massive core-collapse model does manifest in nature (the prediction would be for
SLSNe showing large amounts of radioactive nickel but relatively small amounts of total ejecta). As a final note, it should be
stressed that while the stellar evolution models considered by Yoshida \& Umeda (2011) require stars with exceedingly large initial 
masses ($>310$\,M$_{\odot}$) to form pair-unstable cores at the moderate metallicity indicated for SN 2007bi (Young et al. 2010), alternative models
(Langer et al. 2007) predict that stars with much lower initial masses ($150-250$\,M$_{\odot}$) explode as pair-instability SNe at 
SMC- or LMC-like metallicities, though they may have to be tweaked to
explain the lack of hydrogen in observed SLSN-R spectra. 

  
\subsection{Rates of SLSNe}

The only measurement of the rate of SLSNe is a rough estimate provided by Quimby et al. (2011) based on TSS statistics, which,
normalizing the rate of SLSN-I at $z\approx 0.3$ relative to that of SNe Ia, yields a rate of $\sim10^{-8}$\,Mpc$^{-3}$\,y$^{-1}$.  
This rate is substantially lower than the rates of core-collapse SNe ($\sim10^{-4}$\,Mpc$^{-3}$\,y$^{-1}$), and is also well
below those of rare sub-classes like broad-line SNe Ic (``hypernovae''; $\sim10^{-5}$\,Mpc$^{-3}$\,y$^{-1}$) or long Gamma-Ray 
Bursts ($>10^{-7}$\,Mpc$^{-3}$\,y$^{-1}$; Podsiadlowski et al. 2004; Guetta \& Della Valle 2007).
Both the reported discovery statistics as well as unpublished PTF counts suggest that the rate of SLSN-II is comparable of larger than that
of SLSN-I, while SLSN-R are rarer by about a factor of five, correcting for their slightly lower peak luminosities. I believe this suggests
that SLSN-R are indeed the rarest type of explosions studied so far, and quite possibly arise from stars that are at the very top of the
IMF.

\subsection{The connection between SLSN-I and broad-line Type Ic SNe}

Pastorello et al. (2010) emphasize the similarity between late-time spectra of SN 2010gx (SLSN-I) and spectra
of broad-line Type Ic SNe (see also Quimby et al. 2011; SOM Fig.~\ref{sn2005ap-SYNOW}). They suggest this may
be indicative of a physical connection between all SLSN-I and SNe Ic. However, there are several physical 
differences between SLSN-I and SNe Ic. 

Detailed modelling of SNe Ic (e.g., Sauer et al. 2006; Mazzali et al. 2007; 2010)
indicate the luminosity is dominated by radioactive $^{56}$Ni decay, and these objects show a correlation between
the peak luminosity and the synthesized $^{56}$Ni mass (e.g., Perets et al. 2010; Drout et al. 2011).  The same
is true for SLSN-R (Gal-Yam et al. 2009), but not for SLSN-I (Quimby et al. 2011) in the sense that the nickel mass required
to power the observed luminous peaks is in conflict with the later evolution of the light curve (e.g., Quimby et al 2011; Pastorello et al. 2010). 
The luminosity of SLSN-I must
therefore come from a different source. 

It is interesting to note in this context a small group of very luminous
broad-line SNe Ic (SN 2007D, Drout et al. 2011; SN 2010ay, Sanders et al. 2011) with peak luminosities 
approaching those of SLSN-R ($-20.6$\,mag and $-20.2$\,mag absolute for SN 2007D and SN2010ay, respectively),
but with significantly less $^{56}$Ni ($\sim1$\,M$_{\odot}$). Other processes may be contributing to the large observed 
peak luminosity of these events (e.g., an internal ``engine''; Sanders et al. 2011). Perhaps these are intermediate events
between the class powered purely by radioactivity (normal SNe Ic and SLSN-R) and SLSN-I for which the contribution 
from $^{56}$Ni is negligible, and which must be powered
by some other process, as discussed above. 

Another important physical distinction between SLSN-I and SNe Ic is the size of the emitting region. As shown by
Quimby et al. (2011), the energy radiated by SLSN-I must have been deposited at large initial radii, $\sim10^{15}$\,cm.
On the other hand, early observations of SNe Ic indicate the progenitor stars had an initial radius an order of magnitude smaller,
($<10^{11}$\,cm; Corsi et al. 2011; Sauer et al. 2006) and both the explosion shock energy and and radioactivity must 
therefore be contained within a much smaller initial radius. 

It thus seems that the observed spectroscopic similarity between SLSN-I (at late time) and broad-line SNe Ic suggests
similar ejecta composition and a large kinetic energy (evident as significant amount of mass at high velocities), but other
physical properties (energy source, physical size) are substantially different and argue for a different physical mechanism
powering these two classes of objects. 

\subsection{Forbidden line emission during the photospheric phase of SLSN-R}

As a final curiosity, some photospheric spectra of SLSN-R (Gal-Yam et al. 2009; Young et al. 2010; Gal-Yam 2011; Fig.~\ref{PISNfig}) show
forbidden line emission, notably [Ca II] and probably also [Mg II]. Such lines are usually only observed during the nebular phase
of SNe, when the ejecta are optically thin - this is clearly not the case here. The superposition of this nebular-like emission on an
underlying photospheric spectrum may hint at a complex geometry of the emitting region (motivating spectropolarimetric studies), 
but no explanations for this phenomenon have been put forth so far. 

\section{Summary}

During the last dozen years, numerous super-luminous SN events have been discovered and studied. The accumulated data suggest
these can be grouped into three distinct subclasses according to their observational and physical attributes. Radioactively-powered
SLSN-R seem to be the best understood (and rarest) class, while hydrogen-rich SLSN-II and the most luminous hydrogen-poor 
SLSN-I are more common, but the physical origins of the extreme luminosity they emit is not clear at this time. With several ongoing
surveys efficiently detecting additional examples, the amount of information about these objects, their rates and diversity,  
is likely to increase substantially in the coming few years. 

\section{Acknowledgments}

I would like to thank Ofer Yaron; CCCP members Emilio Enriquez, Alicia Soderberg, S. B. Cenko, Douglad Leonard, Derek Fox, Dae-Sik Moon, 
and David Sand; M. Phillips and P. Nugent; E. Chatzopoulos; L. Chomiuk; S. B. Cenko and Robert Quimby for use of data presented here, 
and E. Nakar, P. Mazzali, D. Xu, S. B. Cenko, A. Soderberg, R. Waldman, A. Pastorello, I. Arcavi, A. Howell, R. Quimby, 
E. O. Ofek, A. Drake, S. Smartt, C. Wheller and A. Miller for useful advice. Members of the 
PTF collaboration and in particular J. S. Bloom, S. B. Cenko, M. M. Kasliwal, S. R. Kulkarni, N. M. Law, P. E. Nugent, E. O. Ofek, 
R. M. Quimby, and D. Poznanski are thanked for use of unpublished PTF material. Anonymous referees are thanked for useful and
constructive suggestions and comments. This work was supported by grants from the Israeli Science Foundation, the US-Israel Binatioal Science 
Foundation, the German-Israeli Foundation, the Minerva foundation, an ARCHES award, and the Lord Sieff of Brimpton fund. 
This research has made use of the NASA/IPAC Extragalactic Database (NED) which is operated by the Jet Propulsion Laboratory, California Institute of Technology, under contract with the National Aeronautics and Space Administration.
 
\newpage 

\centerline{\Large{\bf Supplementary Online Material}}

\section{Defining the threshold cut for superluminous events}

Richardson et al. (2002) conducted a thorough study on the luminosity of supernovae (SNe)
discovered until mid-2001. The most luminous among the common types of SNe are Type Ia
events; Richardson et al. therefore identified a ``SN ridgeline'' at absolute blue magnitude
M$_{B}=-19.5$\,mag (f$=1.2 \times 10^{43}$\,erg\,s$^{-1}$; equal to the peak luminosity of a normal SN Ia) as the 
location where the peak luminosities of cataloged events cluster, and defined ``overluminous
SNe'' as those being significantly more luminous than this value. 

To avoid confusion due to possible errors in measured peak luminosities (arising, e.g., from
distance uncertainties to nearby events), and due to the high-luminosity tail of the scatter around 
the ridgeline value, it is useful to set the arbitrary threshold defining truly super-luminous
supernovae (SLSNe) well away from the ridgeline; examining figure 2 of Richardson et al. (2002) the value
of M$_{B}=-21$\,mag (f$=7.6 \times 10^{43}$\,erg\,s$^{-1}$) seems appropriate. Since the objects
discussed below were discovered over a large redshift range and peak luminosities are reported
in a variety of observed bands, we consider below SNe with reported peak magnitudes
M$<-21$\,mag {\it in any band} as being superluminous. This is justified since the typical bolometric
or cross-filter corrections for SNe around peak are below 1\,mag, and our fiducial threshold is
well above the SN ridgeline. It would probably be better to use the time-integrated luminosity (the
total radiated energy) rather than the peak value as an observational manifestation of unusually
powerful explosions; however, the total energy is more difficult to measure and is unknown for some
objects. Samples of events coming from future surveys may be better observed and allow the use of this
alternative criterion. 

SLSN events were not, so far, detected as luminous sources in wavelengths other than the optical-UV; for example,
see non-detection reports by Gezari et al. (2009) or Chomiuk et al. (2012). This is not surprising in view of the 
distance to most SLSNe discovered so far. The nearby SLSN-II event SN 2006gy is an interesting exception as its
proximity makes detection of, e.g., radio or X-ray emission feasible. Ofek et al. (2007) report non-detection of this
object both in radio and in X-rays. In contrast, Smith et al. (2007) claimed a weak X-ray detection of this event. 
In my opinion, the analysis of Ofek et al. (2007) as well as recent theoretical considerations (Chevalier \& Irwin 2012;
Svirski et al. 2012) suggest this event was not detected in X-rays.
   
\section{Historical review}

\subsection{Cosmological surveys: contaminants}

We now know that superluminous SNe are intrinsically rare. Their discovery therefore
requires surveys that cover a large volume. Such surveys have been conducted in the 
past two decades. The first generation of surveys covering large volumes were designed
to find numerous distant Type Ia SNe for cosmological use. They therefore observed 
relatively small fields of view to a great depth, placing most of the effective survey volume 
at high redshift. These cosmological surveys indeed found mostly high-redshift SNe Ia
(the most luminous type of common SNe) but they also discovered some ``contaminants'' - other
types of luminous explosions at high redshifts (e.g., SN 1995av, Deustua et al. 1995; SN 2000ei, 
Schmidt et al. 2000, See also Neill et al. 2011). 
No published investigations of these objects beyond the initial reports exist, but
they may well have been examples of the super-luminous events we discuss below. 

A more recent 
similar narrow-deep cosmological survey using the Hubble Space Telescope ({\it HST}) indeed
uncovered one of the first bona-fide super-luminous SNe (SLSN-I; SCP 06F6, Barbary et al. 2009). 
Another recent survey, the Canada-France-Hawaii Supernova Legacy Survey (CFH SNLS),
also uncovered unusually luminous events. Among those was the first example of the ``super-chandra''
Type Ia events (SNLS-03D3bb; Howell et al. 2006), a class of SNe Ia that are substantially more luminous
than the SN Ia ridgeline, though not super-luminous according to the criteria defined above. In addition,
the SNLS also detected events that, in retrospect, were very high-redshift ($z\sim1.5$) members of the SLSN-I class
(Howell et al. 2010). 

Finally, re-inspection of data collected as part of the SDSS-II cosmological SN survey uncovered another fine example of 
an SLSN-I (Leloudas et al. 2012); SDSS photometry captures the rising light curve of this object and reveals 
an initial short plateau that may reflect the breaking of the initial explosion shock from a dense envelope.

\subsection{SN factory and the first non-targeted wide-field surveys}

An alternative method to survey a large volume is to use wide-field facilities
to cover a large sky area with relatively shallow imaging. In this way, most of the survey 
volume is at low redshift, and one can conduct an efficient untargeted survey for 
nearby SNe. Such surveys were initially more difficult to conduct since wide-field CCD cameras
were challenging and expensive to construct (while the narrow-deep surveys could be carried out
with smaller focal-plane arrays); analyzing the large data volumes from shallow-wide surveys
was also a computational challenge. However, shallow-wide surveys were eventually undertaken
and provided the first well-observed examples of super-luminous SNe. 

The cosmology group at Lawrence Berkeley Labs pioneered large-scale shallow-wide surveys, initially during the
``1999 spring campaign'' (Regnault et al. 2001) and later on running the ``SN factory'' project (Copin et al. 2006), which
used the Palomar 48'' Schmidt telescope. These surveys were designed to study SNe Ia, but
the fact that the discovered targets were relatively bright allowed high S/N spectroscopy and other
follow-up data to be collected, leading to the discovery of the first relatively well-studied 
superluminous events. The first notable event was SN 1999as (Knop et al. 1999), which turned out to be
the first example of the class of extremely $^{56}$Ni-rich SLSN-R (Gal-Yam et al. 2009; Fig.~\ref{PISNfig}). During the same
campaign, the extremely luminous SN 1999bd was discovered (Nugent et al. 1999; Fig.~\ref{sn1999bd-spec}), which is probably
the first well-documented case of a superluminous SN IIn.  

Interestingly, the Mount Stromlo Abell Cluster Supernova Survey (MSACSS; Reiss et al. 1998) a similar wide-field shallow survey carried out at the southern hemisphere around the same time discovered SN 1997cy (Germany et al. 2000) the other luminous event considered by Richardson et al. (2002) to be 
a bona-fide superluminous event. This event was technically a Type IIn event, but arguments about its real nature persist 
(e.g., Hamuy et al. 2003; Benetti et al. 2006).
Regardless, the absolute peak magnitude of this event (-20.1 mag) as well as those of similar objects discovered since
(e.g., SN 2002ic, Hamuy et al. 2003) fall below our fiducial limit defined above, and we do not discuss them any further.

\subsection{Breakthrough: the Texas Supernova Survey}

Perhaps the most significant breakthrough in our observational knowledge about the most luminous SN explosions resulted from the
operation of the Texas Supernova Survey (TSS) by Quimby and collaborators (Quimby 2006). 
This survey was unique in that the survey telescope, the 0.45m ROTSE IIIb
instrument at McDonald Observatory, had a small aperture, and thus a shallow survey depth, typically 18.5 mag, while the large aperture of 
the main follow-up telescope, the 9.2m Hobby-Eberley Telescope (HET) at the same observatory, enabled high S/N spectroscopic follow-up. While
the number of ROTSE discoveries was modest, it had a high spectroscopic completeness, and in particular did not select against either ``hostless''
events (transients without a visible host galaxy) or events located at galactic nuclei. Previous surveys, which typically suffered from a shortage 
in spectroscopic resources, preferred transients clearly offset from well-resolved galaxies to minimize the chances to observe variable-star or
active galactic nuclei (AGN) contaminants. This search strategy led to several important new discoveries. 

\begin{figure}[h]
\centering
\includegraphics[width=1\textwidth]{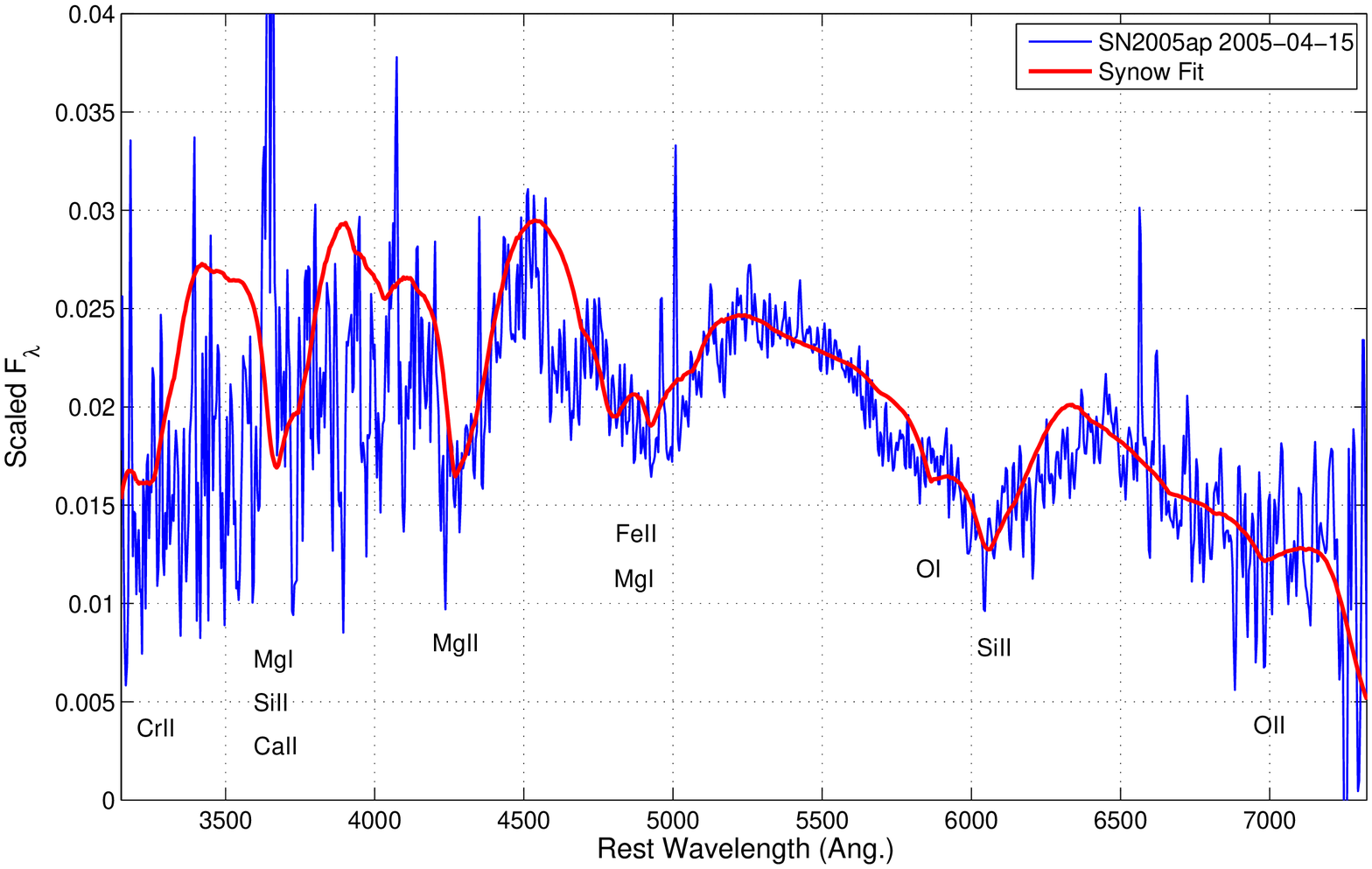}
\caption{A late spectrum of SN 2005ap obtained $\sim36$\,days after peak magnitude (peak date: March 10.6 2005 UT; quimby et al. 2007)
compared to a synthetic SYNOW fit. The spectral features are
well explained by common elements (oxygen, magnesium, silicon and iron); the flux decrement on the blue probably 
results from blends of iron-group elements. The spectrum is overall similar to that of a broad-line Type Ic supernova
(the best fits using the {\it superfit} software (Howell et al. 2005) are of SN 1998bw and SN 2002ap), as pointed out by Pastorello et al. (2010)
for SN2010gx/PTF10cwr. The feature at rest wavelength $6100$\,\AA\, is well-explained by a blend of Si II  and O I lines. 
A contribution from H$\alpha$ previously considered (Quimby et al. 2007) is disfavored since it would require extreme
expansion velocities ($>20000$\,km\,s$^{-1}$); while this was broadly similar to the photospheric velocities seen by
Quimby et al. (2007) in early spectra, it is inconsistent with the moderate photospheric velocity we see in this later 
spectrum ($\sim15000$\,km\,s$^{-1}$), and would further require the hydrogen velocity to increase with time, which
seems unphysical. 
The spectrum (available in digital form from the Weizmann Institute of Science Experimental Astrophysics Spectroscopy System (WISEASS) at http://www.weizmann.ac.il/astrophysics/wiseass/) was obtained using the Double-Beam Spectrograph (DBSP, Oke \& Gunn 1982) 
mounted on the Palomar Observatory 5m Hale telescope on April 15, 2005 UT, as part of the CCCP project (Gal-Yam
et al. 2007b).
}
\label{sn2005ap-SYNOW}
\end{figure}

\begin{figure}[h]
\centering
\includegraphics[width=1\textwidth]{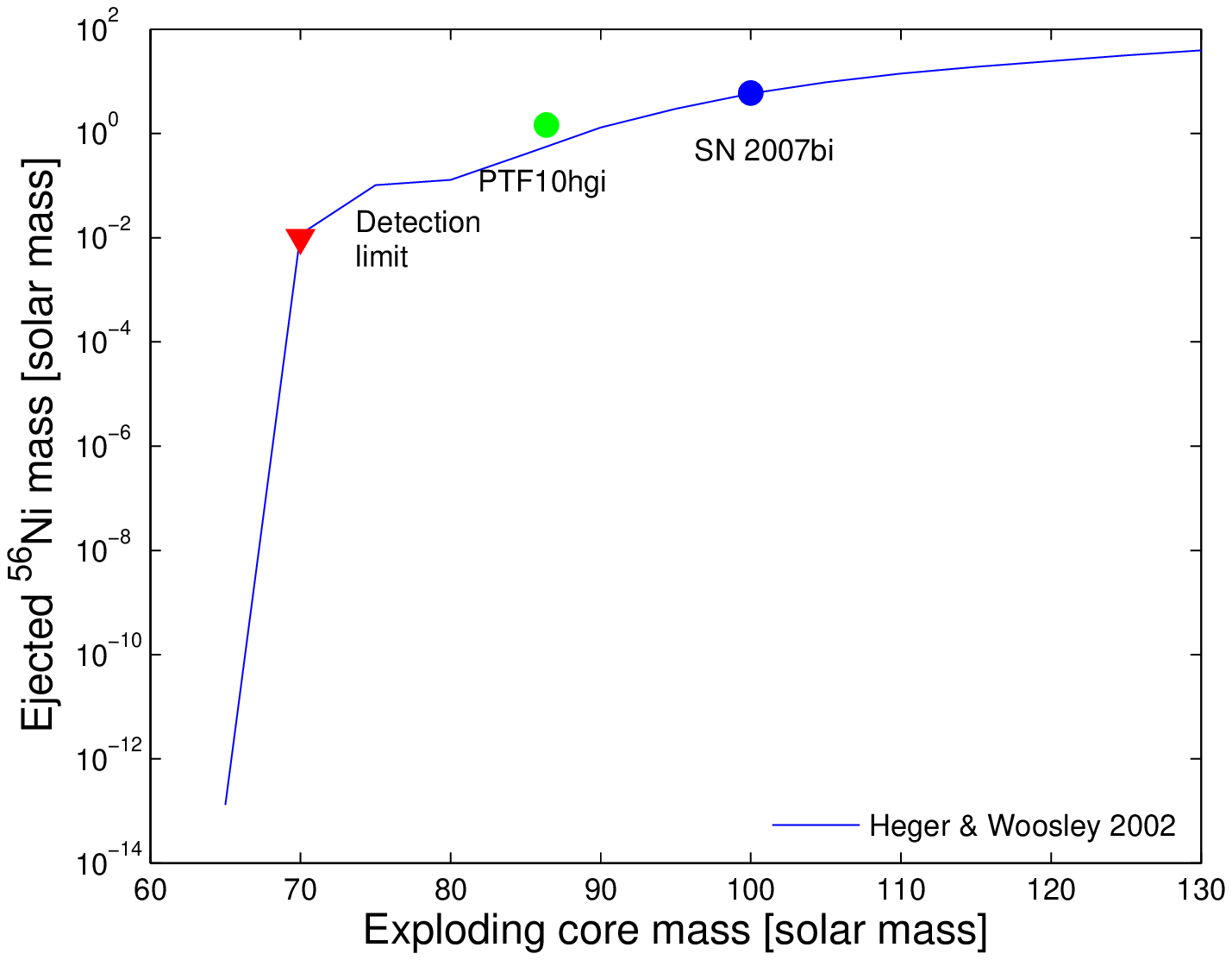}
\caption{The relation between the synthesized $^{56}$Ni mass and the He-core mass (which should roughly equal the 
total ejected mass for Type I events) based on the (non-rotating and non-magnetic) models of 
Heger \& Woosley (2002). Superposed are the data points for SN 2007bi (Gal-Yam et al. 2009) and PTF10hgi (previously unpublished). 
We also mark the approximate upper limit on object detectability (M$_{^{56}{\rm Ni}}=0.01$\,M$_{\odot}$) below which these 
events are expected to be too faint to be discovered anywhere except for the nearest galaxies, where the expected rate of these
events is probably prohibitively low.   
}
\label{PISN-stats}
\end{figure}

On 2005 March 3 this survey discovered a hostless transient at mag 18.13; 
follow-up spectroscopy with HET and the 10m Keck telescope at Mauna Kea
revealed its redshift was $z=0.2832$, indicating an absolute magnitude at peak around $-22.7$\,mag and making
it the most luminous SN discovered until that time (Quimby et al. 2007). 
The interpretation of broad SN features detected was initially difficult (since this was the
first event of its kind) and was limited by the availability of only two spectra covering a short temporal baseline, with limited coverage in the 
red. Retrospective analysis informed by additional events (Quimby et al. 2011), as well as additional spectroscopic data we present here
(Fig.~\ref{sn2005ap-SYNOW}) show that early spectra can be explained almost exclusively by high-excitation lines of intermediate-mass elements
(mostly OII; Quimby et al. 2011), while later spectra evolve to broadly resemble Type Ic supernova spectra of the broad-line subclass 
(Fig.~\ref{sn2005ap-SYNOW}; see also Pastorello et al. 2010); thus SN 2005ap is the first published example of a SLSN-I.   

On 2006 September 18 TSS discovered a bright transient located at the nuclear region of the nearby galaxy NGC 1260 (SN 2006gy; Smith et al. 2007). 
The close proximity of this event to the galactic nucleus (300pc; e.g., Ofek et al. 2007) initially led to suspicion that it was an outburst related to the active 
galactic nucleus, but high angular resolution imaging confirming the slight offset from the galaxy center eventually indicated its supernova nature.
Events of this nature would have probably been missed by most surveys as likely related to AGN; the TSS inclusive approach, as well as the 
fortunate proximity of this event, allowed its recognition as a stellar explosion. 
SN 2006gy suffered substantial host extinction; correcting for this effect, the measured peak magnitude was extreme ($\sim-22$\,mag; Ofek et al. 2007; Smith et al. 2007). Spectroscopy of SN 2006gy 
clearly showed hydrogen emission lines with both narrow and intermediate-width components, leading
to a spectroscopic classification of SN IIn. The initial studies by Ofek et al. (2007) and Smith et al. (2007) were followed by numerous additional 
studies (e.g., Woosley et al. 2007, Agnoletto et al. 2009; Kawabata et al. 2009; Smith et al. 2010, Miller et al. 2010), 
and this object became the prototype and best-studied example of the SLSN-II class.   
TSS went on to discover additional luminous SNe, including the luminous SNe IIn 2006tf (Smith et al. 2008; $\sim-21$\,mag at peak) and
SN 2008am (Chatzopoulos et al. 2011; $-22.3$\,mag at peak) and the unique luminous Type II SN 2008es (Gezari et al. 2009; Miller et al. 2009).

\subsection{Current surveys}

During the last few years, several untargeted surveys have been operating in parallel, and now discover and report most SN events  (Gal-Yam \& Mazzali 2010).
The large volume probed by these surveys and their coverage of a multitude of low-luminosity dwarf galaxies have led, as expected (Young et al.
2008) to the discovery of numerous unusual SNe not seen before in targeted surveys of luminous hosts; indeed, it was shown that the SN population
in dwarf galaxies is different than that observed in giant hosts (Arcavi et al. 2010) . Luminous SNe generally tend
to prefer dwarf hosts (e.g., Neill et al. 2011) and indeed numerous such events are being discovered by this new generation of untargeted 
surveys.
    
The Catalina Real-Time Transient Survey (CRTS) is employing telescopes in Arizona and Australia to search for optical transients
(Drake et al. 2009). As this survey utilizes a catalog-based search method (comparing object catalogs from new
imaging with reference catalogs) it is more sensitive to transients that either have no visible host galaxy, or reside in a 
host that is much fainter than the transient. This makes this survey biased in favor of discovering very luminous SNe, and
indeed many such discoveries have been announced. Prominent results from this survey include the discovery of SN 2008fz,
an extremely luminous member of the SNLS-II family (Drake et al. 2010), as well as of the Type IIn SN 2008iy (Miller et al. 2010)
which did not achieve a bright peak magnitude, but its unprecedented extended light curve implies a large integrated luminosity,
suggesting it may be related to the SNLS-II group. 
 
The Palomar Transient Factory (PTF; Law et al. 2009, Rau et al. 2009) is a wide-field variability survey using the 48'' Oschin Schmidt
Telescope at Palomar Observatory and the refurbished CFHT 12k camera. 
The survey employs image subtraction and is thus sensitive to transients regardless of their host
luminosity. The PTF has been running since 2009, and has discovered and spectroscopically confirmed over 1300 
SNe, including numerous luminous events. The first significant result from PTF was obtained during a ``dry-run'' initial phase of the
project in 2007 conducted using the SN factory infrastructure. This was the discovery of SN 2007bi (Gal-Yam et al. 2009), 
the first extensively-observed example of  the radioactively-powered superluminous SN variety (SLSN-R). 
The PTF discovery of several members of the class of SLSN-I allowed Quimby et al. (2011) to
connect the previously disjoint objects SN 2005ap (Quimby et al. 2007) and SCP 06F6 (Barbary et al. 2009), define the
class of SLSN-I, and derive its main properties from the accumulated data. 

The Panoramic Survey Telescope and Rapid Response System 1 (Pan STARRS 1; hereafter PS1, Kaiser et al. 2010) project employs
a new 1.8m telescope at Haleakala Observatory to conduct wide-field sky surveys. So far, results have been reported mainly from
the medium-deep survey, which covers relatively narrow sky areas to deep limits (i.e., most of its survey volume is at high redshift),
using image subtraction techniques. This appears to be an efficient program to detect numerous SLSNe at high redshifts (out to
$z\sim1$ and beyond); initial results include discovery of two high-redshift SLSN-I (Chomiuk et al. 2011), as well as analysis
emphasizing the potential relation between SLSN-I and less luminous SN Ic varieties (Pastorello et al. 2010).  

\section{Sources of data presented in Figure 1}

The data presented in Figure 1 are take from the following sources. The light curve of SLSN-I PTF09cnd is taken from Quimby et al. (2011).
The data are not corrected for host galaxy extinction, and the distance modulus is calculated from the SN redshift $z=0.258$, assuming the standard
$\Lambda$CDM cosmology with H$_{0}=71$\,km\,s$^{-1}$\,Mpc$^{-1}$. 
$R$-band observations of SLSN-II SN 2006gy are taken from Smith et al. (2007) and Agnoletto
et al. (2009), are corrected for extinction assuming $A_R=1.68$\,mag (Smith et al. 2007) and assuming the distance to NGC 1260 from
the NASA Extragalactic Database (NED; $76.65$\,Mpc, distance modulus of $34.42$\,mag). 
$r$-band observations of SN 2007bi are taken from Gal-Yam et al. (2009), are not corrected for
host galaxy extinction, and use a distance modulus derived from the host galaxy redshift $z=0.1289$ assuming the same cosmology as above.
$R$-band observations of Type IIn SN 2005cl are taken from Kiewe et al. (2011), assuming the host extinction estimated by these 
authors, and the distance modulus to the host galaxy provided by NED. The Type Ia SN light curve is a template $R$-band light curve of a normal
SN Ia provided by Nugent (\url{http://supernova.lbl.gov/$\sim$nugent/nugent$\_$templates.html}). The Type Ib/c light curve is the ``average'' extinction-corrected light curve from Drout et al. 2011. $R$-band Observations of Type IIb SN 2011dh are taken from Arcavi et al. (2011); these are not corrected for host 
galaxy extinction (which these authors 
argue is negligible) and assume the distance modulus to M51 reported there. The light curve of the prototypical SN 1999em is taken from 
Leonard et al. 2002, adjusted to match the absolute plateau luminosity reported by these authors (M$_R=-15.9$\,mag).

\end{document}